\documentstyle[12pt]{article}
\renewcommand{\thepage}{\arabic{page}}
\newcommand{\nc}{\newcommand}
\nc{\beq}{\begin{equation}} \nc{\eeq}{\end{equation}}
\nc{\beqa}{\begin{eqnarray}} \nc{\eeqa}{\end{eqnarray}}
\nc{\lsim}{\begin{array}{c}\,\sim\vspace{-21pt}\\< \end{array}}
\nc{\gsim}{\begin{array}{c}\sim\vspace{-21pt}\\> \end{array}}

\newcommand{\drawsquare}[2]{\hbox{%
\rule{#2pt}{#1pt}\hskip-#2pt
\rule{#1pt}{#2pt}\hskip-#1pt
\rule[#1pt]{#1pt}{#2pt}}\rule[#1pt]{#2pt}{#2pt}\hskip-#2pt
\rule{#2pt}{#1pt}}

\newcommand{\Yfund}{\raisebox{-.5pt}{\drawsquare{6.5}{0.4}}}

\newcounter{mysection}
\newcounter{mysubsection}
\newcommand{\mysection}[1]{\stepcounter{mysection}\setcounter{equation}{0}
\setcounter{mysubsection}{0}\par\bigskip\noindent{\large\bf
\themysection .\ #1}\nopagebreak[4]\par\vskip .3cm}

\newcommand{\mysubsection}[1]{\stepcounter{mysubsection}
\par\medskip\noindent{\large\it
\themysection .\themysubsection\ #1}\nopagebreak[4]\par\vskip .3cm}    
\newcommand{\mysectionstar}[1]{
\par\bigskip\noindent{\large\bf #1}\nopagebreak[4]\par\vskip .3cm}
\def\NF{{N_f}}

\def\Fbar{{\overline{F}}}
\def\l:{\mathopen{:}\,}
\def\r:{\,\mathclose{:}}

\def\inbar{\,\vrule height1.5ex width.4pt depth0pt}
\font\cmss=cmss12 \font\cmsss=cmss12 at 7pt
\def\IZ{\relax\ifmmode\mathchoice
{\hbox{\cmss Z\kern-.4em Z}}{\hbox{\cmss Z\kern-.4em Z}}
{\lower.9pt\hbox{\cmsss Z\kern-.4em Z}}
{\lower1.2pt\hbox{\cmsss Z\kern-.4em Z}}\else{\cmss Z\kern-.4em
Z}\fi}
\def\IB{\relax{\rm I\kern-.18em B}}
\def\IC{{\relax\hbox{$\inbar\kern-.3em{\rm C}$}}}
\def\ID{\relax{\rm I\kern-.18em D}}
\def\IE{\relax{\rm I\kern-.18em E}}
\def\IF{\relax{\rm I\kern-.18em F}}
\def\IG{\relax\hbox{$\inbar\kern-.3em{\rm G}$}}
\def\IP{\relax{\rm I\kern-.18em P}}

\begin{document}

\begin{titlepage}

{\hbox to\hsize{hep-ph/9708134 \hfill EFI-97-35}}
{\hbox to\hsize{August 1997 \hfill Fermilab-Pub-97/295-T}}
{\hbox to\hsize{\hfill UCSD/PTH-97-21}}
\bigskip

\begin{center}

\vspace{.5cm}

\bigskip

\bigskip

\bigskip

{\Large \bf  Chiral Gauge Theories from D-Branes }

\bigskip

\bigskip

{\bf Joseph Lykken}$^{\bf a}$, 
{\bf Erich Poppitz}$^{\bf b,c}$, and {\bf Sandip P. Trivedi}$^{\bf a}$ \\

\bigskip

\bigskip

$^{\bf a}${ \small \it Fermi National Accelerator Laboratory\\
  P.O.Box 500\\
 Batavia, IL 60510, USA\\}

\bigskip

$^{\bf b}${\small \it Enrico Fermi Institute\\
 University of Chicago\\
 5640 S. Ellis Avenue\\
 Chicago, IL 60637, USA\\}

\bigskip

$^{\bf c}${\small \it Department of Physics\footnote{Present address.} \\
University of California at San Diego\\
9500 Gilman Drive\\
La Jolla, CA 92093, USA}
 
\bigskip

\bigskip

{\bf Abstract}
\end{center}
We construct brane configurations leading to chiral four dimensional
$N=1$ supersymmetric gauge theories. The brane realizations consist 
of intersecting Neveu-Schwarz five-branes
and Dirichlet four-branes in non-flat 
spacetime backgrounds. We discuss in some detail the construction 
in a $\IC^2/\IZ_M$ orbifold background.
The infrared theory on the four-brane worldvolume 
is a four dimensional $N=1$ $SU(N)^M$ gauge theory 
with chiral matter representations. We discuss various consistency 
checks and show that the spectral curves 
describing the Coulomb phase of the theory can be obtained once 
the orbifold brane construction is embedded in M-theory. We also 
discuss the addition of extra vectorlike matter and other 
interesting generalizations.

\end{titlepage}

\renewcommand{\thepage}{\arabic{page}}

\setcounter{page}{1}

\mysection{Introduction.}
Recently, the study of Dirichlet branes has led to important insights into
the  behavior of supersymmetric gauge theories. One approach, which has 
proved especially powerful, is to consider  configurations consisting of 
intersecting  Neveu-Schwarz 5-branes  and Dirichlet-branes
\cite{HW}-\cite{CS}.
It was shown by Witten, 
\cite{witten2}, that such  configurations often correspond to a single
$5$-brane in $M$ theory.  A   simple scaling argument  shows that the 
quantum behavior of the resulting gauge theory can then be understood as a 
classical effect in $M$ theory.    So far, in this approach, the background
spacetime before adding branes has been taken to be flat   
(for another  important approach which
considers branes in Calabi-Yau backgrounds see \cite{klemm} and references 
therein), and
the resulting gauge theories have been non-chiral (see, 
however, refs.~\cite{BDL}, \cite{HZ}).  The main purpose of this
paper is to note that 
brane configurations in non-trivial backgrounds can  often lead to 
chiral gauge theories. We illustrate this   by considering brane
configurations consisting of NS 5-branes and intersecting D4-branes in 
a simple class of orbifold backgrounds. As in the flat space case, the 
brane construction  allows us to deduce various features about the 
non-perturbative behavior of these theories. 

This paper is organized as follows. In Section 2, we describe the 
$\IC^2/\IZ_M$ orbifold background
and brane configuration consisting of Dirichlet 
$4$-branes placed at the orbifold point and 
stretched between two Neveu-Schwarz
$5$-branes. The low-energy dynamics is shown to be described by a 3+1 
dimensional $N=1$ theory with $SU(N)^M$ gauge group and  chiral matter
content. In fact,
the gauge theory turns out to be closely related (apart from some 
anomalous $U(1)$ factors)
to the theories
studied in~\cite{IS}, \cite{CEFS}. In Section 3, we study the classical moduli
space of this gauge theory and show that it corresponds to the set of 
allowed motions for the brane configuration; this provides additional 
evidence that we have identified the correct gauge theory.  In Section
4, we turn to the quantum theory and show how by considering the configuration
in $M$ theory one can deduce various
non-perturbative features of the low-energy
dynamics, pertaining to the Seiberg-Witten spectral curves. Finally, some
generalizations of the basic brane configuration are discussed in Section 5. 

This paper is intended to be  a first 
step in a more complete  analysis. Two  further generalizations
are  obvious and will be considered in a subsequent paper. One is to
consider  orientifold backgrounds. The resulting chiral theories are in
many ways more interesting. Another is to  blow up the orbifold and consider
the  brane configuration in the corresponding   ALE space. The resulting smooth
background allows for  a more controlled analysis in $M$ theory. The methods
outlined in this paper give rise to theories which are, in a sense,
closely related to $N=2$ theories. As will become clear below,
their matter content can be thought of as arising from adjoint fields 
after a suitable truncation. These methods might consequently have limited use 
in the study of chiral theories with spinor matter.

\mysection{Brane Configuration and Matter Content. }
\mysubsection{The orbifold and brane configuration.} 
 In this paper we will 
consider $\IC^2/\IZ_M$ orbifolds. 
We choose coordinates so that the $\IC^2$  involved in the 
orbifold  corresponds to the $X^4+ i X^5$ and $X^8+iX^9$ 
directions. The 
Type IIA brane configuration we consider involves two  NS $5$-branes
and several Dirichlet $4$-branes, as shown in Fig. 1. The  NS branes   
stretch along $X^1,X^2,X^3,X^4, X^5$, are placed at the orbifold point,
$X^8=X^9=0$, and  have definite positions in $X^6, X^7$. We take
them to be separated by a finite distance in the 
$X^6$ direction and to be coincident in the $X^7$ direction.  The D4-branes 
are  taken to lie along $X^1,X^2,X^3$, and
$X^6$ directions and end on the two NS branes. 

\mysubsection{The gauge group and  matter content}
 As is well known, the low-energy dynamics of this configuration is described 
by a 3+1 dimensional
 field theory, which lives in the intersection region of the D$4$ 
branes and NS branes. We will show below that 
 $NM$ $4$-branes placed at the origin of the $\IZ_M$ 
orbifold give rise to an $N=1$ $U(N)^M$ gauge theory. The  matter content
consists of chiral superfields which transform under the gauge groups as:
\begin{equation}
\label{mattercontent}
 \begin{array}{c|ccccc}
      & U(N)_1 & U(N)_2 & U(N)_3 & \cdots & U(N)_M \\ \hline
  Q_1 & \Yfund  & \overline{\Yfund}  & 1       & \cdots & 1 \\
  Q_2 & 1       & \Yfund  & \overline{\Yfund}  & \cdots & 1 \\
  \vdots & \vdots & \vdots & \vdots & \vdots & \vdots \\
  Q_M & \overline{\Yfund} & 1 & 1 & \cdots & \Yfund 
 \end{array} \nonumber
\end{equation}
Note that the matter  content is chiral\footnote{
One overall $U(1)$ factor above is ``frozen out" while the remaining $U(1)$s
are anomalous; we will have more to say on this below.}. 

We now turn to justifying this 
claim  for the gauge group and matter content.  
First consider the number of supersymmetries. In the absence of the 
orbifold this brane configuration preserves 8
supercharges or  $N=2$  supersymmetry  in 3+1 dimensions: the IIA
theory has $32$ supercharges,
but the presence of 4-branes and NS branes reduces that by
a factor of $2 \cdot 2$.
In the $\IZ_M$ orbifold we only keep gravitino states for
which the  vertex  operators  are invariant under a rotation by 
$\exp \{ ({2 \pi i \over M }) (J_{45}-J_{89}) \}$. This further reduces 
the supersymmetry by half leading  to $4$ supercharges or $N=1$ in 
3+1 dimensions. 

To arrive at the gauge group and matter content it is useful to consider
the final configuration built up in two stages. Let us first look at a
configuration without the NS branes where the 4-branes are infinite along
$X^6$ and  are placed at the orbifold point. It is well known
for compact orbifolds that
tadpoles must cancel in the one-loop vacuum amplitude,
and that   this constraint is often powerful enough to determine the
gauge group and matter content \cite{GP}.  In our case, 
the one loop amplitude only
receives a contribution from the cylinder diagram and is easy to work out.  
Since the $\IC^2$ on which the orbifold group acts is noncompact,
we do not expect any constraint on the allowed  total number of 4-branes:
the corresponding RR flux can always escape to
infinity. This is borne out by an explicit calculation. However, there are 
non-trivial constraints which arise from the tadpole cancellation for  twisted
RR fields. Let the orbifold group $\IZ_M$ act 
on $v = X^4 + i X^5$ and $w = X^8 + i X^9$ as:
\beq
\label{ZMaction}
 ( v,~ w )
\rightarrow (\alpha ~v, ~ \alpha^{-1} ~w)~, ~~ \alpha \equiv 
e^{2 \pi i\over M}~~,
\eeq
and  the corresponding action of the orbifold group on the Chan Paton 
factors $\lambda$ be represented by a matrix $\gamma_{\alpha}$:
\beq
\label{egamma}
\lambda \rightarrow \gamma_\alpha ~ \lambda ~
\gamma_\alpha^{-1}~.
\eeq 
The $4$-branes are sources of twisted RR
scalars that can only propagate 
 in one of the directions transverse to the 4-branes ($X^7$). 
As argued in \cite{BI}, a one-volume is insufficient to allow
the Ramond-Ramond flux to escape to infinity, and 
the tadpole cancellation condition must be satisfied even for infinite volume.
The constraints from the twisted RR tadpoles are then given by:
\beq
\label{tadpoleconditions}
{\rm tr} ~\gamma_\alpha^K ~=~0, ~~ K = 1,\ldots ,M-1~.
\eeq
Note that $\gamma_{\alpha}$ must furnish a 
representation of the orbifold group and thus $\gamma_{\alpha}^M=1$. This
together with eq. (\ref{tadpoleconditions}) allows us to solve for
$\gamma_{\alpha}$. We find, first, that the number of $4$-branes at the 
orbifold point must be a multiple of $M$; we refer to this number  hereafter
as $N M$. Second, we find that the matrix $\gamma_{\alpha}$, in a suitable
basis, is given by:
\beq
\label{solutionforgamma}
\gamma_\alpha ~=~{\rm diag} \{1 \times {\bf 1}_N, \alpha 
\times {\bf 1}_N,\ldots , \alpha^{M-1} \times {\bf 1}_N\} ~,
\eeq
with ${\bf 1}_N$ being the unit $N\times N$ matrix.
The gauge and matter content can now be worked out as well. 
The corresponding gauge group on the $4$-brane worldvolume theory
turns out to be $U(N)^M$. 
Fluctuations in the $X^7$ direction which survive the orbifold projection
contribute one adjoint field for each $U(N)$ factor. Together with the gauge
bosons these form an $N=1$ vector multiplet in 4+1 dimensions.  Finally,
from the $X^4,X^5,X^8,X^9$ directions we get hypermultiplets transforming 
under the gauge groups as described in  eq.~(\ref{mattercontent}) (we note
that the same orbifold has been considered in \cite{DM}, \cite{BI}).

\begin{figure}[h]
\begin{center}
\setlength{\unitlength}{0.240900pt}
\ifx\plotpoint\undefined\newsavebox{\plotpoint}\fi
\sbox{\plotpoint}{\rule[-0.200pt]{0.400pt}{0.400pt}}%
\begin{picture}(1500,900)(0,0)
\font\gnuplot=cmr10 at 10pt
\gnuplot
\sbox{\plotpoint}{\rule[-0.200pt]{0.400pt}{0.400pt}}%
\put(275,917){\makebox(0,0){{\bf NS}}}
\put(939,917){\makebox(0,0){{\bf NS}}}
\put(508,270){\makebox(0,0)[l]{{\bf D6}}}
\put(342,594){\makebox(0,0)[l]{{\bf D4}}}
\put(1270,1079){\makebox(0,0){{\bf 4,5}}}
\put(1187,715){\makebox(0,0){{\bf 7,8,9}}}
\put(1469,877){\makebox(0,0)[l]{{\bf 6}}}
\put(1270,877){\vector(1,0){166}}
\put(1270,877){\vector(0,1){162}}
\multiput(1268.92,874.16)(-0.499,-0.729){163}{\rule{0.120pt}{0.683pt}}
\multiput(1269.17,875.58)(-83.000,-119.582){2}{\rule{0.400pt}{0.342pt}}
\put(1187,756){\vector(-2,-3){0}}
\sbox{\plotpoint}{\rule[-0.600pt]{1.200pt}{1.200pt}}%
\put(275,68){\usebox{\plotpoint}}
\put(275.0,68.0){\rule[-0.600pt]{1.200pt}{194.888pt}}
\put(939,68){\usebox{\plotpoint}}
\put(939.0,68.0){\rule[-0.600pt]{1.200pt}{194.888pt}}
\sbox{\plotpoint}{\rule[-0.200pt]{0.400pt}{0.400pt}}%
\put(275,473){\usebox{\plotpoint}}
\put(275.0,473.0){\rule[-0.200pt]{159.958pt}{0.400pt}}
\put(275,513){\usebox{\plotpoint}}
\put(275.0,513.0){\rule[-0.200pt]{159.958pt}{0.400pt}}
\put(275,553){\usebox{\plotpoint}}
\put(275.0,553.0){\rule[-0.200pt]{159.958pt}{0.400pt}}
\put(275,432){\usebox{\plotpoint}}
\put(275.0,432.0){\rule[-0.200pt]{159.958pt}{0.400pt}}
\put(275,392){\usebox{\plotpoint}}
\put(275.0,392.0){\rule[-0.200pt]{159.958pt}{0.400pt}}
\sbox{\plotpoint}{\rule[-0.500pt]{1.000pt}{1.000pt}}%
\put(508,311){\usebox{\plotpoint}}
\multiput(508,311)(10.887,17.671){19}{\usebox{\plotpoint}}
\put(707,634){\usebox{\plotpoint}}
\put(458,311){\usebox{\plotpoint}}
\multiput(458,311)(10.887,17.671){19}{\usebox{\plotpoint}}
\put(657,634){\usebox{\plotpoint}}
\put(408,311){\usebox{\plotpoint}}
\multiput(408,311)(10.887,17.671){19}{\usebox{\plotpoint}}
\put(607,634){\usebox{\plotpoint}}
\put(557,311){\usebox{\plotpoint}}
\multiput(557,311)(10.887,17.671){19}{\usebox{\plotpoint}}
\put(756,634){\usebox{\plotpoint}}
\end{picture}
\end{center}
\caption{The Type IIA brane configuration for the
$\IZ_M$ orbifold models. There are $N$ physical four-branes
stretched between the NS five branes, plus their
$\IZ_M$ images. The six branes,
if present, give extra vectorlike matter. The bending
of the NS branes is not shown in this figure.}
\end{figure}
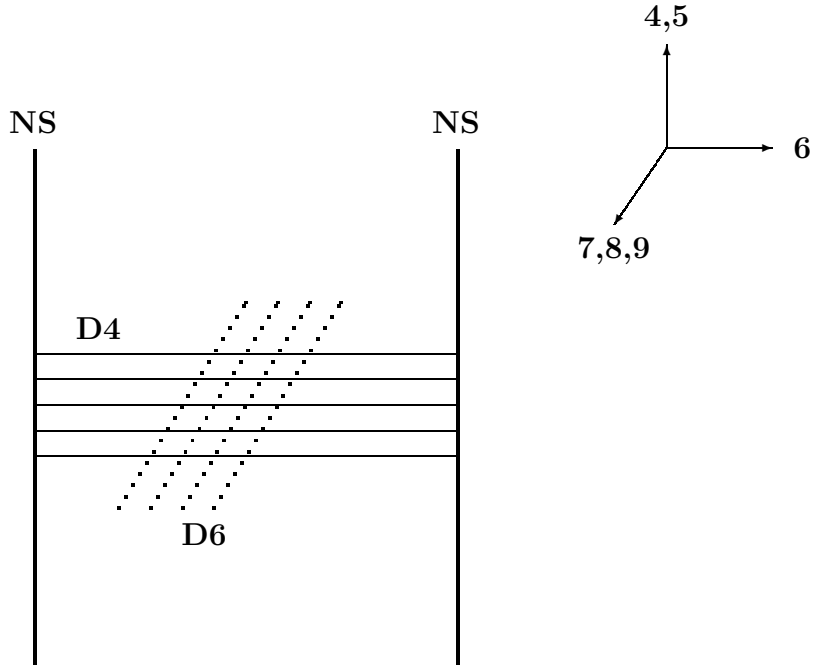

Now finally we can add the two NS branes and sandwich the four-branes between
them as in Fig. 1. What is the resulting 3+1 dimensional theory? 
It is useful for this purpose to describe the above matter content in the
language of 3+1 dimensions.  The  component of the gauge field, $A_6$, can be
paired with the  adjoint fields coming
from the $X^7$ direction to give a chiral
superfield. Each hypermultiplet  will transform as two chiral
multiplets in $3+1$ dim. language, one of the two chiral multiplets coming from
fluctuations in the $X^4, X^5$ directions, and the other from the  $X^8,X^9$ 
directions. One expects the boundary conditions coming from the ends of the 4
brane, where it terminates on the 5-brane, to  freeze some of these degrees 
of freedom. Based on the analysis in the absence of the orbifold one expects
the gauge field to survive  and  the chiral mulitiplet coming from the
$(A_6, X^7)$, fluctuations to be  frozen. Similarly,
the matter coming from the 
fluctuations in the $X^4, X^5$ directions should survive whereas that from the 
$X^8, X^9$ directions will be frozen out. This finally gives rise to the 
$U(N)^M$ theory with the matter content described in eq.~(\ref{mattercontent}).
We note again that each field in eq.~(\ref{mattercontent})
represents a chiral multiplet so that the theory is  chiral.

Above, we  first considered the $4$-branes without NS branes  in the orbifold
background and then introduced the  NS branes. It is also 
illuminating to consider things  in the opposite order.  Accordingly, let us
first consider a configuration of $NM$ $4$-branes  stretched between the two NS
branes in the absence of the  orbifold.
The resulting field theory is well known
to be an $N=2$ theory, with $SU(NM)$ gauge group. The adjoint scalar field
corresponds to fluctuations of the $4$ branes along 
the $X^4, X^5$ directions. It is
natural to expect that  the orbifold should correspond to implementing a
projection in this theory. In fact, the gauge theory possesses a 
$U(1)$ global  symmetry under
which (in $N=1$ language) the gauge field
and its fermionic partner transform as
$(A_{\mu}, \lambda) \rightarrow  (A_{\mu}, \lambda)$,
  and  the adjoint and its fermionic
partner as $(\phi, \psi) \rightarrow e^{i \alpha} (\phi, \psi)$. 
 In general this  symmetry is anomalous, however
it has a non-anomalous $\IZ_{2 N M}$ discrete  subgroup. This discrete subgroup
in turn has a $\IZ_M$ subgroup.   In
addition, the gauge symmetry has a $\IZ_M$ discrete subgroup  under which a
fundamental representation is multiplied by  
$ {\rm diag} \{1 \times {\bf 1}_N,
\alpha \times {\bf 1}_N,\ldots ,  \alpha^{M-1} \times {\bf 1}_N\} $, with 
$ {\bf 1}_N$ being the unit $N\times N$ matrix.
In the $N=2$ field theory  it is
natural  to identify the orbifold group with the product  of these two $\IZ_M$
symmetries.  On doing so and retaining states invariant under this product
 discrete symmetry one gets  precisely the $U(N)^M$ group and matter content 
mentioned above. 

\mysection{Brane Motion and the Classical Moduli Space.}
In this section we  compare the   set of
allowed motions of the brane configuration to the classical moduli space of the
gauge theory described above. This will serve two purposes. First, agreement
between the two will give  additional evidence  that we have identified
the  correct gauge theory. Second, in the process we will understand better
the  role of the various $U(1)$s in this theory---an issue which we have so 
far not fully addressed. 

It will be convenient in the following discussion to organize the  
$U(1)$s in the following basis. We will choose the first $U(1)$  to be the 
sum  of the $U(1)$ factors, and the other $U(1)$s to be orthogonal 
to the first. It is easy to see from eq.~(\ref{mattercontent})
that none of the matter 
fields are  charged under the first $U(1)$.   In fact one can deduce that  
this $U(1)$ factor is frozen, i.e. its coupling vanishes. There are two
arguments in support of this. First, for the case of a flat space time 
background, it was argued in \cite{witten2}, that in the  $N=2$ theory 
this overall $U(1)$ must be frozen.  We saw above that for the orbifold
background the resulting  field theory could be understood as a further
truncation of the $N=2$ theory; we  thus expect the $U(1)$ to continue to 
be frozen in it. Second, we will see below that when we interpret this 
configuration in $M$ theory, the genus of the two dimensional 
surface spanned by
the  5-brane worldvolume will be  
consistent with the absence of the $U(1)$.  

Turning our attention to the remaining $U(1)$s we notice that they are all
anomalous\footnote{The $\IZ_2$ orbifold is an exception:
in this case the theory 
is not chiral.}. These $U(1)$s are analogous to anomalous $U(1)$ factors
which often arise in string compactifications \cite{DSW}. 
In the context of D-branes
anomalous $U(1)$s were discussed in \cite{DM} where they were shown to play 
an important role in governing the low-energy dynamics. We will discuss these
$U(1)$s in some detail below. Here we summarize their essential features
which are important in the present discussion of the classical moduli space. 
The important point is that these anomalous $U(1)$s are  broken. 
The low-energy 3+1 dimensional theory contains axion fields which arise 
from twisted RR fields,
and the anomalies are cancelled by shifting these axions
appropriately \cite{GS}. In fact the axions can be regarded as the longitudinal
components of the heavy gauge bosons. 
The only feature that  is really  important  in the present discussion is that 
each $U(1)$ will give a D-term  contribution to the full 
potential energy, which is important in determining the moduli space 
of the theory (notice also that the $U(1)$ charges are all traceless, hence no 
Fayet-Iliopoulos
term is generated by loop effects).  

We are now ready to study the motion of the 4-branes.  We begin with  a
$\IZ_M$ orbifold with   $NM$ branes located at the orbifold point.  
The corresponding gauge group is $SU(N)^M$.  The 
$4$-branes  can only move along the $X^4, X^5$ directions,
 since  they end on NS branes
which  only extend along these directions.   Each  4-brane  has $M-1$ images 
under the $\IZ_M$ symmetry,  so counting images, 
we can move  sets of $M$ branes away from the orbifold point. 
Moving $M$ branes away breaks $SU(N)^M \rightarrow SU(N-1)^M
\times U(1)$.  If all the $4$-branes are moved away from the orbifold point we
are left with  a  $U(1)^{N-1}$ gauge symmetry. Since the motion of each set 
of $M$ branes is described by one complex number, the moduli space is $N$ 
dimensional. Finally, we also note that if $N_1$ physical branes come
together away from the orbifold point we get an enhanced $U(N_1)$ gauge 
symmetry. 

Now consider the flat directions in the gauge theory. These 
are in one-to-one correspondence with gauge invariant chiral superfields
made out of the elementary matter fields in  eq.~(\ref{mattercontent}). 
Ignoring the anomalous $U(1)$s for the moment, these moduli are
of two kinds.  One class is best described in terms of the operator:
\beq
\label{stmp}
\Sigma^i_j =(Q_1 \cdot \ Q_2 \cdots \ Q_M )^i_j,
\eeq
as:
\beq
\label{fltdireca}
\phi_k =tr(\Sigma)^k,
\eeq
for $k=1, \cdots , (N-1) $.
The second class of ``baryonic" directions is given by:
\beq
\label{flatdirecb}
b_{\alpha}=(Q_{\alpha})^N,
\eeq
with  $\alpha = 1, \cdots , M $. 
Altogether, we see that there are $N-1+M$ flat directions; these are more 
than the number of brane degrees of freedom found above.  
The discrepancy is corrected when we account for the $D$-term potential 
generated by the anomalous $U(1)$s. 
We saw above that there are $M-1$ of these,
thus their $D$ terms get rid of $M-1$ moduli giving us, finally, a $N$ 
dimensional moduli space in agreement with what we found for the motion of 
branes.
An analysis of the vacuum expectation values also  shows that in 
the moduli space, 
generically, a $U(1)^{(N-1)}$ is left unbroken.  
Finally, one finds subspaces of the moduli space
which correspond to partially enhanced gauge symmetry, again in accord with
what is found from brane considerations. 

\mysection{The Quantum Behavior via $M$ Theory}

We will now turn to considering the quantum behavior of the gauge theory
described above. It was found in the previous section that generically 
in moduli space the theory is in the Coulomb phase with the 
gauge symmetry being broken to a $U(1)^{(N-1)}$ subgroup. 
We would like to see if the corresponding spectral curves,
\cite{seibergwitten},
can be determined. In this analysis we will closely follow \cite{witten2}
where it was pointed out that in $M$ theory,   the brane configuration 
corresponding to that in Fig. 1 can be thought of as the worldvolume of a 
single NS 5-brane, and that this insight leads to determining the curves. 

In \cite{witten2} the 5-brane worldvolume had infinite extent along 
the $X^0,X^1,X^2,X^3$ coordinates, while spanning a two dimensional surface in
the four-manifold parametrized by
$v=X^4+iX^5$ and $t=\exp (-s)=\exp (-(X^6+iX^{10})/R)$.
In our case $v$ and $w=X^8+iX^9$ are modded by the $\IZ_M$
transformation eq.~(\ref{ZMaction}). A more convenient representation
of this $\IC^2/\IZ_M$ orbifold is obtained by embedding it as
a hypersurface in $\IC^3$:
\beq
\label{orbsurf}
yz - x^M = 0.
\eeq
The coordinate mapping is $y=v^M$, $z=w^M$, $x=vw$; the orbifold
singularity is at $y=z=x=0$. In the $M$ theory limit the 5-brane
is described by a Riemann surface $\Sigma$ embedded in
$\IC^3 \times R^1 \times S^1$. 
This surface is smooth except
at the orbifold point, and can be parametrized as a rational curve
by $y$ and $t$, with $z$ set equal to zero.

Now consider the configuration shown in Fig. 1, consisting of two NS branes
and $NM$ $4$-branes (we are counting the branes and their images as distinct)
 stretching between them. The two dimensional surface 
$\Sigma$ can now be described
by the curve:
\beq
\label{swcurve}
t^2 + B(y) ~t +1=0.
\eeq
Here $B$ is a polynomial of degree $N$ (in $y=v^M$), i.e.,
\beq
\label{defb}
B(y)= y^N+ u_1 y^{N-1} + u_2 y^{N-2} + \cdots + u_N.   
\eeq
Note this surface corresponds to  genus $N$$-$$1$ 
as would be expected 
for a curve with $N$$-$$1$ photons. As discussed in \cite{witten2} and 
\cite{seibergwitten}, the periods of this Riemann surface determine the gauge 
couplings of the $N$$-$$1$  $U(1)$ gauge groups.

The asymptotic behavior of $t$ for large $y$ is given by 
$t \simeq - y^N$, and $t\simeq -y^{-N}$. This tells us how the two 
NS branes bend for large $y$ and determines the  asymptotic form of the 
beta function which goes like 
\beq
\label{bfun}
{4 \pi \over g^2} \simeq 2N \ln |y| ~.
\eeq
This agrees with the expected beta function for each of the $SU(N)$ factors.

The coefficients $u_i$ in eq.~(\ref{swcurve}) parametrize the moduli space
of the theory. It would be useful to express them in terms of the gauge 
invariants built out of the elementary fields in eq.~(\ref{mattercontent}). 
When the 4-branes are sufficiently far (compared to the strong coupling 
scale(s)) from the orbifold point the leading order dependence of the 
 $u_i$ can be determined by 
classical considerations. To see this, note that eq.~(\ref{swcurve}), at  fixed
 $t$, can be used to solve for $y$ and thereby yield the positions of the 
4-branes. Furthermore, at large enough separation these positions can 
be unambiguously related to the gauge invariants, thereby determining the 
leading dependence of the $u_i$.

In Section 2, we had described how the gauge theory corresponding to $NM$ 
$4$-branes placed at a $\IZ_M$ orbifold point can be thought of 
as being obtained by starting from an $SU(NM)$, $N=2$, theory and only keeping 
states invariant under a certain $\IZ_M$ symmetry. In fact this provides
the simplest way of determining the leading dependence of the coefficients 
$u_i$. One starts with the $N=2$ curve,
\beq
\label{swntwo}
t^2+B(v)~t+1=0,
\eeq  
where $B(v)$ is a polynomial of degree $NM$ given by:
\beq
\label{defbntwo}
B(v)=v^{NM} + a_1 v^{NM-1} +a_2 v^{NM-2} + \cdots + a_{NM}.
\eeq
In this case the coefficients are easily determined as (trace of) 
the appropriate
powers of the adjoint field. 
We now only allow fields
invariant under the $\IZ_M$ symmetry to have vacuum expectation values. 
This means
that only integer powers of $v^M$ survive in $B(v)$. 
The resulting curve thus has a $\IZ_M$ symmetry, under which 
$v \rightarrow e^{{2 \pi i \over M}} v$. 
To obtain the curve in the orbifold theory it is natural to identify points 
related by this symmetry. 
 This amounts to parametrizing the curve with 
a variable $y= v^M$. The curve, eq.~(\ref{swntwo}), then turns into the 
required one, eq.~(\ref{swcurve}). As mentioned before, the coefficients
in eq.~(\ref{swntwo}) can be determined in terms of the adjoint field and 
can then be easily expressed in terms of the moduli in the orbifold theory.

The leading dependence of the $u_i$ on the gauge invariants can thus 
be determined. However, there can be subleading terms in these relations, 
depending on 
strong coupling scales of the gauge theories involved, which 
cannot be 
determined by classical considerations alone\footnote{Such 
terms are absent in the $N=2$ $SU(N_c)$ theory 
studied in \cite{witten2}, provided $N_c>N_f$.
In this case 
dimensional arguments and the fact that these corrections arise from instanton
effects and are therefore proportional to $\Lambda^{b_0}$ is enough to 
explain their absence.}.
In fact, such terms are present in the theories at hand. 
We know this because these theories are essentially identical, (apart from the 
anomalous $U(1)$s discussed above) to the $SU(N)^M$ theories studied
in \cite{IS}, \cite{CEFS}, 
and  their curves have been worked out from field theoretic considerations.

For illustrative purposes we consider the example of an $SU(2)^3$ theory,
which corresponds to taking six $4$-branes (two physical branes and their
images) in  a $\IZ_3$ orbifold.
The related theory was discussed in  \cite{CEFS}
and the curve was obtained to be:
\beq
\label{swft}
{\tilde t}^2=(x^2-(\Lambda_1^4 \ M_2+ \Lambda_2^4 \ M_3 + \Lambda_3^4 \ M_1
               -M_1 M_2 M_3 +T^2))^2 - 4 \Lambda_1^4 \Lambda_2^4 \Lambda_3^4.
\eeq
Here, $\Lambda_{1,2,3}$ are the three strong coupling scales, while
$M_i=Q_i^2$ and $T \sim Q_1\cdot Q_2 \cdot Q_3$ are the moduli. 
This curve is related to 
the one obtained in the brane construction, eq.~(\ref{swcurve}) by a shift 
and rescaling of the variables $y$ and $t$. 
On doing so and comparing one finds that 
the $M_1 M_2 M_3$ and $T^2$ terms
in the first bracket in eq.~(\ref{swft}) correspond to the leading 
dependence of the coefficients $u_i$, while the $\Lambda$-dependent terms
in the first bracket correspond to the subleading terms we were worried
about. 
Actually, strictly speaking we need to incorporate the effects of the 
 anomalous $U(1)$s in the curve, eq. (\ref{swft}), before comparing the two.
 This is relatively simple to do 
in the orbifold limit where the Fayet-Iliopoulos terms for the two anomalous
$U(1)$s are zero\footnote{Determining the  curve away from the orbifold 
limit is an interesting problem which we hope to address in a subsequent paper.
This will also allow us to see whether the subleading terms 
arise in part because 
of the orbifold nature of the background and can be determined by blowing it
up.}.   

Let us pause for a moment to sketch this out. In a convenient basis,
the $U(1)$ charge assignments of the three elementary  fields are,
 $Q_1 : (2,0)$, $Q_2:(-1,1)$, $Q_3: (-1,-1)$. 
The corresponding $D$ terms then imply:
\beq
\label{dterma}
4|M_1|^2-2|M_2|^2-2|M_3|^2=0,    
\eeq
and
\beq
\label{dtermb} 
2|M_2|^2-2|M_3|^2 =0.
\eeq
In this example,  the $U(1)$ anomalies  cancel due to appropriate shifts
in two  axion fields. One consequence is that the $\Lambda$ dependent terms in 
eq.~(\ref{swft}) acquire an axion  dependence.  
In describing the resulting curve it is simplest to carry out 
appropriate $U(1)$ rotations (and shifts in  the axion fields) to go 
to a gauge where the three fields $M_1$, $M_2$ and $M_3$ have the same 
phase.  Eq.~(\ref{dterma})  and (\ref{dtermb})  can now be used to solve 
for two of the fields,
say, $M_2$ and $M_3$ in terms of third,  $M_1$. On substituting back in 
eq.~(\ref{swft}) the  resulting curve in this gauge in terms of the 
moduli, $M_1$ and $T$ is 
given by:
\beq
\label{finalcurve}
{\tilde t}^2=(x^2-(\Lambda_1^4+ \Lambda_2^4  + \Lambda_3^4) \ M_1
               +M_1^3 -T^2)^2 - 4 \Lambda_1^4 \Lambda_2^4 \Lambda_3^4.
\eeq 
The axion dependence in eq.~(\ref{finalcurve})
 enters through the dependence of the strong coupling scales
on these fields and can be easily worked out. 
 The important point is that
after going through this procedure, in eq.~(\ref{finalcurve})
one sees that the subleading terms mentioned above continue to persist, 
while the leading terms on which there was agreement in the two cases are
not changed in an essential way.

\mysection{Generalizations of the Orbifold Brane Construction.}
\mysubsection{Additional vectorlike matter.}
There are three obvious generalizations of our $\IZ_M$
orbifold brane construction which add massless vectorlike matter.

The first 
is obtained by adding $M \cdot N_f$ Dirichlet six branes at the origin in the
$(X^4,X^5)$ plane. These 6-branes extend in the directions
$X^1, X^2, X^3, X^7, X^8, X^9$ and do not
break any additional supersymmetries \cite{david}. Once again,
the tadpoles
in the one-loop vacuum amplitude must cancel in this theory.
The only additional  constraints arise from the 
 twisted Ramond-Ramond tadpole amplitudes for strings ending
on these 6-branes\footnote{The twisted RR amplitudes from 4-6 strings
can be already seen to vanish since the matrices representing
the action of the twist on the Chan-Paton factor of the 4-branes
obey (\ref{tadpoleconditions}).}: the 6-branes are sources of twisted
RR flux which can only propagate in the $X^6$ transverse direction, which
is insufficient to allow the flux to escape to infinity \cite{BI}. Therefore,
 the total 
twisted RR charge of the 6-branes has  to vanish, and 
 the matrices that  represent the $\IZ_M$ action on the 
six brane Chan-Paton factors must also obey the conditions
(\ref{tadpoleconditions}). 

The massless excitations of the $4-6$ strings give  
vectorlike matter with the following
transformation law under the $SU(N)^M \times SU(N_f)^M$ 
symmetry (here we have denoted by $SU(N_f)$ the $6$-brane
gauge group, which appears as a global symmetry in the
$4$-brane theory). There are $M$ fields $F_i$ ($i = 1, \cdots , M$),
transforming
as $(\Yfund, {\bar{\Yfund}})$ under $SU(N)_i \times SU(N_f)_i$, which
are singlets under the other gauge and flavor groups, and $M$ fields
$\bar{F}_i$ that transform as $({\bar{\Yfund}}, \Yfund)$ under
$SU(N)_i \times SU(N_f)_{(i + 1)({\rm mod} M)}$ (and, similarly, are 
singlets under the other gauge and flavor groups). The shift of 
indices for the $\bar{F}$ fields is due to the fact that the vertex
operator for the massless $4-6$ string excitations transforms by
a factor of $e^{-i \pi/M}$ under the $\IZ_M$ symmetry
\cite{GP} and  the Chan-Paton factors for the 6-branes obey
$\gamma_\alpha^M = -1$. Finally, as a vestige of $N=2$ supersymmetry, 
the following Yukawa couplings that preserve the $SU(N_f)^M$ global
symmetry 
will appear in the superpotential:
\beq
\label{superpotential}
W ~ = ~ F_1Q_M\Fbar_M + F_2Q_1\Fbar_1 + F_3Q_2\Fbar_2 + \cdots
 + F_MQ_{M-1}\Fbar_{M-1}~.
\eeq
It will become clear in the following that this 
spectrum (and superpotential) is the only one consistent with
field theoretic considerations and nonabelian duality.

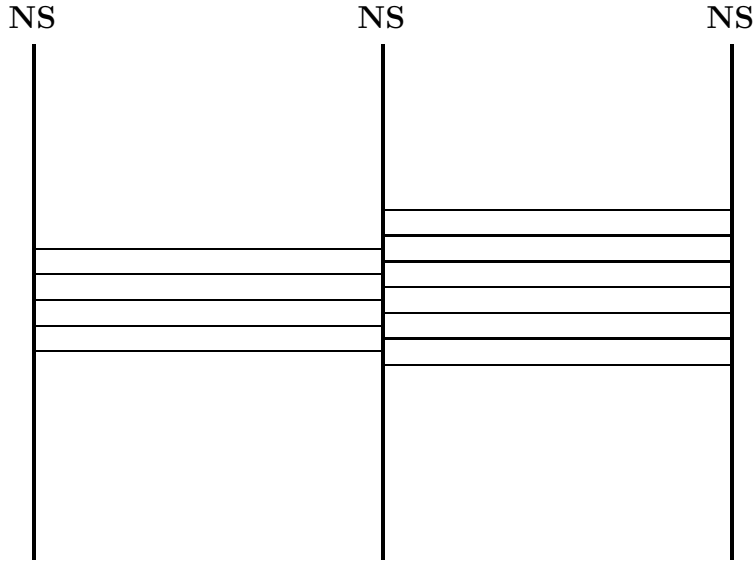
\begin{figure}[h]
\begin{center}
\setlength{\unitlength}{0.240900pt}
\ifx\plotpoint\undefined\newsavebox{\plotpoint}\fi
\sbox{\plotpoint}{\rule[-0.200pt]{0.400pt}{0.400pt}}%
\begin{picture}(1500,900)(0,0)
\font\gnuplot=cmr10 at 10pt
\gnuplot
\sbox{\plotpoint}{\rule[-0.200pt]{0.400pt}{0.400pt}}%
\put(258,917){\makebox(0,0){{\bf NS}}}
\put(806,917){\makebox(0,0){{\bf NS}}}
\put(1354,917){\makebox(0,0){{\bf NS}}}
\sbox{\plotpoint}{\rule[-0.600pt]{1.200pt}{1.200pt}}%
\put(258,68){\usebox{\plotpoint}}
\put(258.0,68.0){\rule[-0.600pt]{1.200pt}{194.888pt}}
\put(806,68){\usebox{\plotpoint}}
\put(806.0,68.0){\rule[-0.600pt]{1.200pt}{194.888pt}}
\sbox{\plotpoint}{\rule[-0.200pt]{0.400pt}{0.400pt}}%
\put(258,473){\usebox{\plotpoint}}
\put(258.0,473.0){\rule[-0.200pt]{132.013pt}{0.400pt}}
\put(258,513){\usebox{\plotpoint}}
\put(258.0,513.0){\rule[-0.200pt]{132.013pt}{0.400pt}}
\put(258,553){\usebox{\plotpoint}}
\put(258.0,553.0){\rule[-0.200pt]{132.013pt}{0.400pt}}
\put(258,432){\usebox{\plotpoint}}
\put(258.0,432.0){\rule[-0.200pt]{132.013pt}{0.400pt}}
\put(258,392){\usebox{\plotpoint}}
\put(258.0,392.0){\rule[-0.200pt]{132.013pt}{0.400pt}}
\put(806,493){\usebox{\plotpoint}}
\put(806.0,493.0){\rule[-0.200pt]{132.013pt}{0.400pt}}
\put(806,533){\usebox{\plotpoint}}
\put(806.0,533.0){\rule[-0.200pt]{132.013pt}{0.400pt}}
\put(806,574){\usebox{\plotpoint}}
\put(806.0,574.0){\rule[-0.200pt]{132.013pt}{0.400pt}}
\put(806,614){\usebox{\plotpoint}}
\put(806.0,614.0){\rule[-0.200pt]{132.013pt}{0.400pt}}
\put(806,452){\usebox{\plotpoint}}
\put(806.0,452.0){\rule[-0.200pt]{132.013pt}{0.400pt}}
\put(806,412){\usebox{\plotpoint}}
\put(806.0,412.0){\rule[-0.200pt]{132.013pt}{0.400pt}}
\put(806,371){\usebox{\plotpoint}}
\put(806.0,371.0){\rule[-0.200pt]{132.013pt}{0.400pt}}
\sbox{\plotpoint}{\rule[-0.600pt]{1.200pt}{1.200pt}}%
\put(1354,68){\usebox{\plotpoint}}
\put(1354.0,68.0){\rule[-0.600pt]{1.200pt}{194.888pt}}
\end{picture}
\end{center}
\caption{The Type IIA brane configuration for a class of
generalized $\IZ_M$ orbifold models. There are $N$ physical four-branes
stretched between the first pair of NS five branes, plus their
$\IZ_M$ images. There are $N_f$ physical four-branes
stretched between the second pair of NS five branes, plus their
$\IZ_M$ images.}
\end{figure}

Another generalization is obtained by adding more NS branes.
The simplest example is illustrated in Fig. 2.
This configuration can be obtained by starting with
$\NF$ physical four-branes stretched between two NS branes 
without any six-branes. 
 One then brings a third
NS brane in from infinity along the $X^7$ direction, until
it intersects the middle of the four-branes. One can then break
the four-branes on this new NS brane; the gauge group at
this point is clearly $SU(\NF)^M \times SU(\NF)^M$.
Now one can move $\NF - N$
of the left-hand physical four-branes together with their
$\IZ_M$ images off to infinity in the $(X^4,X^5)$ plane,
where they have no effect on the light spectrum of
the remaining brane configuration.
Thus we deduce that Fig. 2 represents an orbifold model with gauge group
$SU(N)^M \times SU(\NF)^M$, with chiral matter content
under $SU(N)^M$ and $SU(\NF)^M$ of the form
(\ref{mattercontent}). In addition there is vectorlike
matter corresponding to chiral multiplets
$F_i$ ($i = 1, \cdots , M$),
transforming
as $(\Yfund, {\bar{\Yfund}})$ under $SU(N)_i \times SU(N_f)_i$,
$\bar{F}_i$ that transform as $({\bar{\Yfund}}, \Yfund)$ under
$SU(N)_i \times SU(N_f)_{(i + 1) ({\rm mod} M)}$ (both $F$ and $\bar{F}$ are
singlets under all the other gauge groups).
There is  also a superpotential,
which is the sum of the superpotentials (\ref{superpotential}) for the 
gauge groups $SU(N)^M$ and $SU(N_f)^M$, respectively.

The third generalization consists of attaching semi-infinite
four-branes to the left- or right-hand NS brane. This is
equivalent (modulo the discussion in \cite{SS}) 
to taking the configuration of Fig. 2 and
moving the left- or right-hand NS brane off to infinity
in the $X^6$ direction. The new vectorlike matter consists
of $\NF$ flavors for each $SU(N)$, with superpotential (\ref{superpotential})
 (in the limit that the left- or right-hand NS brane is pushed off to 
infinity, the gauge coupling of the corresponding $4$-brane theory
goes to zero and the contribution to the superpotential from the $SU(N_f)^M$
gauge group vanishes). It appears therefore that this construction is related
to the construction with $6$-branes by the Hanany-Witten process:
after pushing the $6$-branes through one of the NS branes, a set  of 
$N_f$ $4$-branes stretched between the NS brane and the $6$-branes
is created; after moving the $6$ branes to infinity we obtain the construction
with semi-infinite $4$-branes described above.

It is clear that the generalizations discussed above can also be obtained
from  $N=2$ theory with matter after eliminating states that are not invariant
under  an appropriately chosen $\IZ_M$ discrete global symmetry, in the
same way that was discussed in the end of Section 2 for the pure
Yang-Mills $N=2$ theory.

\mysubsection{Nonabelian duality.}
 
Here we discuss one more check on our orbifold construction with
extra matter fields. The $N=1$ 
theory with gauge group $SU(N)^M$, with matter
content given by eq.~(\ref{mattercontent}) plus additional $N_f$ flavors
of each $SU(N)$ factor, and  a superpotential
given by eq.~(\ref{superpotential})
was   considered in ref.~\cite{ES}. By an iterative application of the $N=1$
SQCD dualities it was found that the theory has an equivalent infrared
description---along the Higgs branch---in terms of an $SU(N_f - N)^M$ 
theory with the same matter content and superpotential. The theories
along their respective  Coulomb branches are clearly different, as
follows from the different number of unbroken $U(1)$s at
a generic point on the Coulomb branch moduli space. 

The $SU(N_c)^M \Leftrightarrow SU(N_f - N_c)^M$ duality is 
related to the duality of the Higgs branches of $N=2$ SQCD with gauge
groups $SU(M N_c)$ and $SU(M(N_f - N_c))$. 
This duality is easy to see in the brane construction \cite{HW}, \cite{david}.
Consider the brane configuration of Fig. 1. Pushing the $N_f$
6-branes (we count only the  physical branes here)
to the left of the left NS brane, we obtain a configuration with $N_f$ 
4-branes stretching between the $N_f$ $6$-branes and the left NS
brane. Then we enter the Higgs branch of the theory by reconnecting
the $N_c$ 4-branes stretching between the two NS branes with $N_c$
of the newly created $4$-branes and rearranging them in the most
general way consistent with the $s$-rule \cite{david}. Thus we
obtain a configuration where $N_c$ $4$-branes stretch between 
the $6$-branes and the right NS brane while $N_f - N_c$ 
$4$-branes stretch between the 6- branes and
the left NS brane. Now we can move the two NS branes past 
each other in the $X^6$ direction and reconnect once more 
the 4-branes, obtaining thus a configuration 
where $N_f - N_c$ $4$-branes stretch between the two NS branes, 
and $N_f$ $4$-branes between the 6-branes and the leftmost NS brane. 
This setup describes the Higgs branch moduli space of 
the $SU(M(N_f - N_c))$ $N=2$ theory with $N_f$ flavors.  Orbifolding by 
the $\IZ_M$ symmetry does not affect the previous argument
in any essential way. We thus obtain a brane realization of the Higgs
branch duality between the $SU(N_c)^M$ and $SU(N_f - N_c)^M$ theories.

\mysectionstar{Acknowledgements}
We thank Atish Dabholkar, Michael Douglas, Ken Intriligator and  David 
Kutasov for helpful discussions.
JL acknowledges the hospitality of the Aspen Center for
Physics, where portions of this research were completed.
The research of JL and ST is supported by the Fermi National
Accelerator Laboratory, which is operated by Universities Research
Association, Inc., under contract no. DE-AC02-76CHO3000.
E.P. was supported by DOE contract no. DOE-FG03-97ER40506.

\nc{\ib}[3]{ {\em ibid. }{\bf #1} (19#2) #3}
\nc{\np}[3]{ {\em Nucl.\ Phys. }{\bf #1} (19#2) #3}
\nc{\pl}[3]{ {\em Phys.\ Lett. }{\bf #1} (19#2) #3}
\nc{\pr}[3]{ {\em Phys.\ Rev. }{\bf #1} (19#2) #3}
\nc{\prep}[3]{ {\em Phys.\ Rep. }{\bf #1} (19#2) #3}
\nc{\prl}[3]{ {\em Phys.\ Rev.\ Lett. }{\bf #1} (19#2) #3}


\begin{thebibliography}{99}

\bibitem{HW}A. Hanany and E. Witten, hep-th/9611230, \np{B492}{97}{152}.

\bibitem{ooguri1}J. de Boer, K. Hori, Y. Oz and Z. Yin, hep-th/9702154.

\bibitem{david}S. Elitzur, A. Giveon, and D. Kutasov, 
hep-th/9702104, \pl{B400}{97}{269}; 

S. Elitzur, A. Giveon, D. Kutasov, E. Rabinovici,
and A. Schwimmer, hep-th/9704104.

\bibitem{barbon}J.L.F. Barb\` on, hep-th/9703051,
\pl{B402}{59}{1997}.

\bibitem{witten2}E. Witten, hep-th/9703166.

\bibitem{EJS} N. Evans, C.V. Johnson and A. D. Shapere, hep-th/9703210.

\bibitem{BH}J.H. Brodie and A. Hanany, hep-th/9704043.

\bibitem{T}R. Tatar, hep-th/9704198.

\bibitem{LLL}K. Landsteiner, E. Lopez and D. A. Lowe, hep-th/9705199.

\bibitem{BSTY}A. Brandhuber, J. Sonnenschein, S. Theisen and S. Yankielowicz,
hep-th/9704044; hep-th/9705232.

\bibitem{HZ}A. Hanany and A. Zaffaroni, hep-th/9706047.


\bibitem{ooguri} K. Hori, H. Ooguri and Y. Oz, hep-th/9706082.

\bibitem{W2}E. Witten, hep-th/9706109.

\bibitem{BIKSY}A. Brandhuber, N. Itzhaki, V. Kaplunovsky, J. Sonnenschein,
and S. Yankielowicz, hep-th/9706127.


\bibitem{AT}C. Ahn and R. Tatar, hep-th/9707106; C. Ahn, K. Oh,
 and R. Tatar, hep-th/9707027.

\bibitem{NOS}S. Nam, K. Oh, and S.-J. Sin, hep-th/9707247.

\bibitem{HSZ}A. Hanany, M.J. Strassler and A. Zaffaroni, hep-th/9707244.

\bibitem{NOYY}T. Nakatsu, K. Ohta, T. Yokono, and Y. Yoshida, hep-th/9707258.

\bibitem{SS}M. Schmaltz and R. Sundrum, hep-th/9708015.

\bibitem{CS}C. Cs\` aki and W. Skiba, hep-th/9708082.


\bibitem{klemm} A. Klemm, hep-th/9705131. 

\bibitem{BDL}M. Berkooz, M.R. Douglas and R.G. Leigh, hep-th/9606139,
\np{B480}{96}{265}.

\bibitem{IS}K. Intriligator and N. Seiberg, hep-th/9408155, \np{B431}{94}{551}.


\bibitem{CEFS}C. Cs\' aki, J. Erlich, D. Freedman, and W. Skiba, 
hep-th/9704067.

\bibitem{GP}E. Gimon and J. Polchinski, hep-th/9601038, \pr{D54}{96}{1667}.

\bibitem{BI}J. Blum and K. Intriligator, hep-th/9705030.

\bibitem{DM}M. Douglas and G. Moore, hep-th/9603167.

\bibitem{DSW} M. Dine, N. Seiberg, and E. Witten, \np{B289}{87}{589}.

\bibitem{GS} M. B. Green and J.H. Schwarz, \pl{B149}{84}{117}. 

\bibitem{seibergwitten}N. Seiberg and E. Witten, 
hep-th/9407087, \np{B426}{94}{19}.

\bibitem{ES}N. Evans and M. Schmaltz, hep-th/9609183, \pr{D55}{97}{3776}.

\end{thebibliography}
\end{document}